# An improved limit on $^{171}$Yb electric dipole moment

S.-Z. Wang[1,*] S.-B. Wang[1,*] E. F. Zhou[2,3], Z.-J. Tao[1], J.-M. Yao[2,3,†] T. Xia[4,‡] and Z.-T. Lu[1,4§]
[1] *CAS Center for Excellence in Quantum Information and Quantum Physics, School of Physical Sciences, University of Science and Technology of China, Hefei, Anhui, 230026, China.*
[2] *School of Physics and Astronomy, Sun Yat-sen University, Zhuhai, Guangdong, 519082, China.*
[3] *Guangdong Provincial Key Laboratory of Quantum Metrology and Sensing, Sun Yat-sen University, Zhuhai, Guangdong, 519082, China. and*
[4] *Hefei National Laboratory, University of Science and Technology of China, Hefei 230088, China*

Searches for permanent electric dipole moments (EDMs) probe new sources of CP violation beyond the Standard Model. The atomic EDM of $^{171}$Yb, which is sensitive to its nuclear Schiff moment, is a promising candidate, but previous measurements were limited by noise and systematic effects induced by the optical dipole trap (ODT) that holds the atoms. Here we report a substantially improved search enabled by a dual-ODT configuration with orthogonal polarizations, and spin measurements for precise alignment of static and optical fields. By suppressing the key parity-mixing systematic effect and vector light-shift noise, we achieve a limit of $|d(^{171}\mathrm{Yb})| < 7.5 \times 10^{-28}\ e \cdot \mathrm{cm}$ (95% C.L.), which represents a 20-fold improvement over the previous work. We also present the first calculation of the nuclear structure factors of the Schiff moment of $^{171}$Yb. The experimental and theoretical advances combine to tighten constraints on CP-violating pion-nucleon couplings in a global analysis. Moreover, this work demonstrates a pathway toward future measurements in $^{225}$Ra, where nuclear deformation offers exceptional sensitivity to physics beyond the Standard Model.

*Introduction*—Permanent electric dipole moments (EDMs) provide a sensitive window into new sources of CP violation [1–3]. Within the Standard Model, predicted EDMs are far below current experimental search sensitivities [2]. Indeed, no EDM has been found so far. However, the observed matter–antimatter asymmetry in the Universe strongly suggests that there exist additional CP-violating mechanisms beyond the Standard Model, motivating increasingly precise EDM searches across a variety of systems.

EDM searches in different categories—including lepton EDM [4–6], nucleon EDM [7], and nuclear Schiff moment [8–15]—probe complementary combinations of CP-violating interactions. Studies of the Schiff moments have employed vapor or gas cell ($^{199}$Hg [8, 9], $^{129}$Xe [10, 11]), molecular beam ($^{205}$TlF [12, 16]), and optically trapped atoms ($^{225}$Ra [13, 14], $^{171}$Yb [15]). Within this category, the $^{199}$Hg experiment has achieved the highest search sensitivity, at the $10^{-30}\ e \cdot \mathrm{cm}$ level. Nevertheless, less precise measurements in other nuclides still play an essential role, as their results are combined with $^{199}$Hg in global analyses to place tighter constraints on multiple hadronic CP-violating parameters [17–19].

The diamagnetic atom $^{171}$Yb offers a promising platform for atomic EDM and nuclear Schiff moment studies [20, 21]. Laser-cooled $^{171}$Yb atoms held in an optical dipole trap (ODT) have previously been used to measure the EDM [15]. This approach enables strong electric fields, long coherence times, and suppression of the $\mathbf{v}\times\mathbf{E}$ systematic effect. However, effects due to the strong ODT light imposed limitations on both the statistical and systematic uncertainty, restricting the sensitivity to the $10^{-27}\ e \cdot \mathrm{cm}$ level, insufficient for probing physics beyond the Standard Model.

Here we report a substantially improved search for the EDM of $^{171}$Yb. A dual-ODT configuration with orthogonal polarizations suppresses a key parity-mixing systematic effect in the previous work. Spin measurements provide feedback for precise alignment of both the static and ODT optical fields, further suppressing noise due to vector light shift of the ODT, which is a differential shift between Zeeman sublevels induced by the residual circular polarization component of the ODT field [22]. Combining these advances with a higher electric field, the limit on the $^{171}$Yb Schiff moment is improved by a factor of 20, surpassing the $^{225}$Ra, $^{129}$Xe, and $^{205}$TlF systems. On the theory side, we perform the first nuclear-structure-based calculation of the $^{171}$Yb Schiff moment nuclear structure factors. Together, these experimental and theoretical advances yield stringent constraints on CP-violating pion-nucleon interactions in a global analysis.

*Experimental setup*—$^{171}$Yb atoms are first cooled to 50 $\mu$K in a two-stage magneto-optical trap (MOT), and then transported by a movable ODT to an adjacent science chamber [15]. The EDM measurement part, including both the setup and the procedure, is significantly improved over the prior work. In this work, to suppress the parity-mixing effect, we use two pairs of ODTs: one parallel pair, named $\mathrm{ODT}_{+k,\mathrm{H}}$ and $\mathrm{ODT}_{+k,\mathrm{V}}$, propagates in the $+z$ direction; the counter-propagating pair are named $\mathrm{ODT}_{-k,\mathrm{H}}$ and $\mathrm{ODT}_{-k,\mathrm{V}}$ (Fig. 1(a)). Each pair is generated with a calcite crystal, which splits the output into two parallel ODTs spatially separated by 0.6 mm, and with their linear polarizations in the orthogonal $x$ and $y$ directions (horizontal and vertical, labeled H and V, respectively). The angle between H and V polarizations deviates from $\pi/2$ by less than 0.1 mrad. Both pairs are derived from the same laser at 1035.9 nm, the magic wavelength for the $6s^2\,{}^1S_0 \leftrightarrow 6s6p\,{}^3P_1$ transition [23]. At this wavelength, the AC Stark shifts of the lower and upper states are equal so that the transition frequency remains unchanged. Each ODT has a power of 15 W, a beam waist of 40 $\mu$m, and a trap depth of 200 $\mu$K. Approximately $1 \times 10^6$ atoms are loaded into each ODT, with a trap lifetime of 50 s.

* These authors contributed equally to this work.
† Corresponding author: yaojm8@sysu.edu.cn
‡ Corresponding author: txia@hfnl.cn
§ Corresponding author: ztlu@ustc.edu.cn

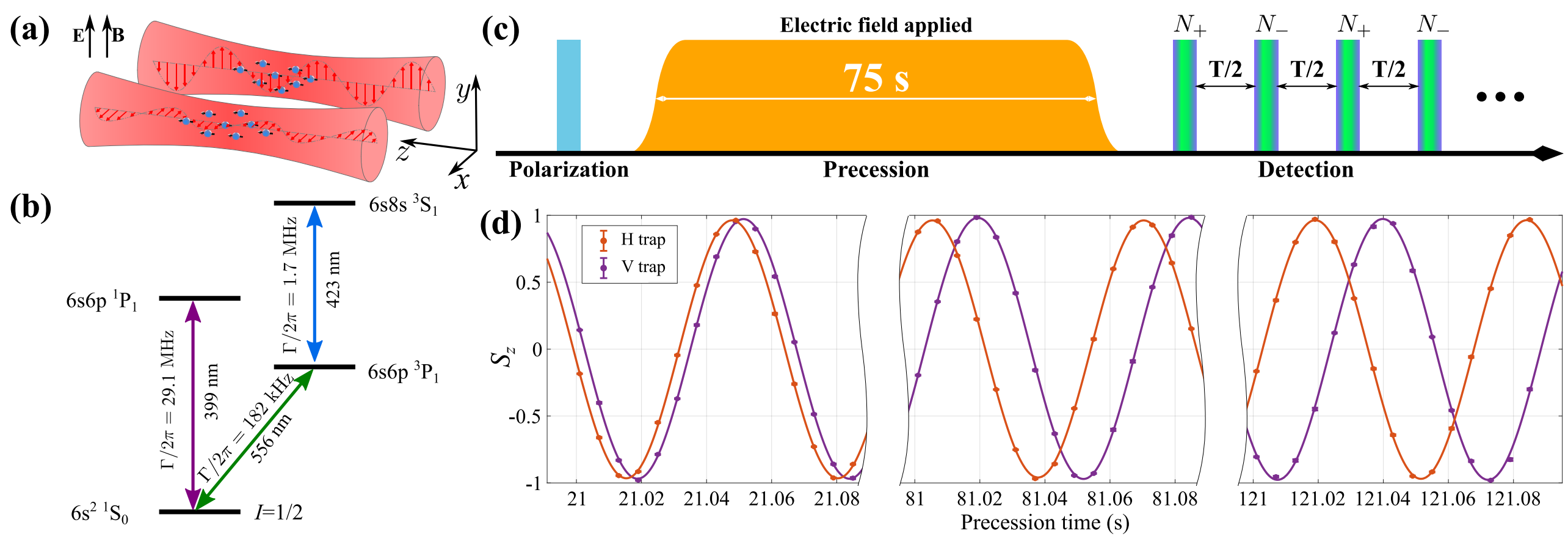

FIG. 1. **(a)** Two pairs of parallel ODTs are used alternately in the EDM measurement. A bias magnetic field of 20 mG along $+y$ drives spin precession at $2\pi \times 15$ Hz. An electric field of 236 kV/cm is applied also along the $y$-direction with alternating polarities. **(b)** Energy levels and transitions for spin polarization ($^1S_0 \leftrightarrow {}^1P_1$, 399 nm), population detection ($^1S_0 \leftrightarrow {}^3P_1$, 556 nm) and level dressing ($^3P_1 \leftrightarrow {}^3S_1$, 423 nm) used in a QND method. **(c)** Measurement sequence. While the B field is constantly on, the E field is applied during the precession stage. The accumulated phase under the combined B and E fields is read out via a QND method. **(d)** Precession signals of the two ensembles in ODT$_{+k,\mathrm{H}}$ (travelling along $+z$, linearly polarized along $x$, labeled H for horizontal, red curve) and ODT$_{+k,\mathrm{V}}$ ($+z$, $y$, V for vertical, purple curve). Their frequencies differ by 0.03% due to a residual $B$-field gradient.

The atoms are positioned by the ODTs between a pair of 10 mm-diameter copper electrodes spaced 0.76 mm apart. The bottom electrode is grounded, while the top one can be ramped to $\pm 18$ kV within 5 s, generating a $y$-directed electric field. The field strength is calibrated to be $\pm 236$ kV/cm by measuring the DC Stark shift of the $6s^2\,{}^1S_0 \leftrightarrow 6s6p\,{}^3P_1$ transition [24] (see Supplemental Material for details). Single-point diamond polishing of the electrode surfaces is key to achieving this stable, strong electric field. Furthermore, a stable and uniform magnetic field is achieved by enclosing the science chamber in a multi-layer magnetic shield with a shielding factor of $\sim 10^5$, and by using a $\cos(\theta)$ coil within the shields to generate a 20 mG bias field along the $y$-direction.

The measurement employs three laser beams (Fig. 1(b)): a 399 nm pump beam, a 423 nm dressing beam, and a 556 nm probe beam. The probe beam and the dressing beam are applied simultaneously in the quantum non-demolition detection (QND) of the spin precession phase [25]. These beams are combined using a dichroic mirror, all circularly polarized ($\sigma^+$) and propagating along the $+z$ direction. With an elliptical beam profile of 0.3 mm $\times$ 3 mm, the beams pass cleanly through the 0.76 mm electrode gap in the vertical ($y$-axis) direction while fully covering the pair of ODTs separated by 0.6 mm horizontally. The atomic population is determined via absorption imaging of the probe beam.

The ground state $6s^2\,{}^1S_0$ of $^{171}$Yb ($I = 1/2$) has two energy sub-levels split by the combined interactions between the magnetic dipole moment ($\mu$) and the $B$ field ($B_s$), and between the EDM ($d$) and the $E$ field ($E_s$): $\hbar\omega_\pm = 2\mu B_s \pm 2dE_s$, where the $+$ ($-$) sign corresponds to the $E$ field being parallel (antiparallel) to the $B$ field. To measure the precession frequency corresponding to the energy splitting (Fig. 1(c)), the spins are first polarized along $z$; then undergo precession around the $y$ axis; finally the spin projection along $z$, $S_z = (N_+ - N_-)/(N_+ + N_-)$, is read out by a QND method [25], where $N_\pm$ is the spin population along $\pm z$. The time evolution follows $S_z = Pe^{-t/\tau}\cos(\omega t)$, where $P$ denotes the spin polarization and $\tau$ the coherence time. Fig. 1(d) shows precession curves for atoms in ODT$_{+k,\mathrm{H}}$ and ODT$_{+k,\mathrm{V}}$, both with $\omega = 2\pi \times 15$ Hz. The precession amplitude remains above 0.97 after 120 s of precession, indicating a much longer coherence time.

A single phase measurement takes 100 s, including 20 s for preparation (mainly loading and transport) and 80 s for precession. Two consecutive measurements—with the $E$ and $B$ fields parallel and antiparallel, respectively—yield phases $\phi_+$ and $\phi_-$ ($\phi_\pm = \omega_\pm t$), based on which the EDM is extracted as $d = \hbar(\phi_+ - \phi_-)/(4E_0 T_{\mathrm{eff}})$, where $T_{\mathrm{eff}} = \int_0^{T_0} [E_s(t)/E_0]\,dt$ accounts for the ramping of the electric field, $T_0$ is the total precession time, and $E_0$ denotes the steady-state field strength.

*Parity-mixing effect*—In EDM measurements on atoms confined in an ODT, the combined interaction of a static electric field $E_s$ (unit vector $\boldsymbol{\epsilon_s}$), a static magnetic field $B_s$ ($\mathbf{b_s}$), and the AC electric ($\boldsymbol{\epsilon}$) and magnetic field ($\mathbf{b}$) of the ODT light induces a shift in the spin-precession frequency [21]:

$$\nu_{\rm PM} = \nu^{(1)}(\mathbf{b}\cdot\mathbf{b_s})(\boldsymbol{\epsilon}\cdot\boldsymbol{\epsilon_s}) + \nu^{(2)}(\mathbf{b}\cdot\boldsymbol{\epsilon_s})(\boldsymbol{\epsilon}\cdot\mathbf{b_s}). \quad (1)$$

Here the two frequency parameters are defined as $\nu^{(1,2)} = IE_s\mathcal{R}^{(1,2)}$, where $I$ is the ODT light intensity, and $\mathcal{R}^{(1,2)}$ are atomic structure coefficients. Notice that the expression in Eq. (1) is both P-even and T-even, but EDM-like since it reverses sign as the static electric field is inverted. This parity-mixing effect generates an EDM systematic effect: $\tilde{d} = h(\nu_{\rm PM,+} - \nu_{\rm PM,-})/4E_s$, where $\nu_{\rm PM,+}$ and $\nu_{\rm PM,-}$ are the frequency shifts induced by the parity-mixing effect under positive and negative high-voltage (HV) polarities, respectively.

The parity-mixing effect was observed by Zheng *et al.* [15]. In this work, we examine the dependence of this effect on multiple experimental parameters (including directions and angles; see Fig. 2) in order to develop an advanced suppression strategy. A single ODT traveling along $+z$ is used in this investigation, for which the parity-mixing EDM systematic effect can be expressed as

$$\tilde{d}_{+k} = \tilde{d}_1\sin(2\theta - \alpha) + \tilde{d}_2\sin(\alpha), \quad (2)$$

where $\theta$ is the angle between $\boldsymbol{\epsilon}$ and $\boldsymbol{\epsilon_s}$ of the negative-HV configuration, $\alpha$ is the angle between $\mathbf{b_s}$ and $\boldsymbol{\epsilon_s}$ (negative-HV), and the coefficients are defined as $\tilde{d}_1 = -hI(\mathcal{R}^{(1)} + \mathcal{R}^{(2)})/4$ and $\tilde{d}_2 = hI(\mathcal{R}^{(1)} - \mathcal{R}^{(2)})/4$. Fig. 2 presents the investigation results under three different configurations. Fitting these data using Eq. (2) yields $\tilde{d}_1 = (-2.0 \pm 0.2) \times 10^{-25}$ $e\cdot$cm and $\tilde{d}_2 = (7.0 \pm 0.2) \times 10^{-25}$ $e\cdot$cm. The ODT intensity experienced by the atoms is independently calibrated by measuring the AC Stark shift of the $6s^2\,{}^1S_0 \leftrightarrow 6s6p\,{}^3P_1$ transition, yielding $I = (1.45 \pm 0.11) \times 10^6$ W/cm$^2$ (see Supplemental Material for details). The corresponding atomic-structure coefficients are then derived to be $\mathcal{R}^{(1)} = (2.40 \pm 0.19) \times 10^{-7}$ a.u. and $\mathcal{R}^{(2)} = (-1.33 \pm 0.13) \times 10^{-7}$ a.u. (a.u.: atomic units). These parameters are not only essential for assessing systematic uncertainties in EDM measurements, but also provide valuable benchmarks for future atomic-structure calculations.

Based on the geometric configuration of the dual ODTs containing the two orthogonal polarizations (H and V), we identify the following cancellation scheme: $\tilde{d}_{+k,\rm V} + \tilde{d}_{+k,\rm H} = 2\tilde{d}_2\sin(\alpha)$. Using a pair of co-propagating ODTs composed of a vertically polarized ODT$_{+k,\rm V}$ and a horizontally polarized ODT$_{+k,\rm H}$, the polarization-mismatch issue is effectively resolved, leaving a residual component depending on the angle $\alpha$ between the static $E$ and $B$ fields. This residual component can be further suppressed by employing another pair of counter-propagating ODTs, through the previously established $k$-vector reversal scheme [15, 21]. The residual parity-mixing systematic uncertainty arising from intensity imbalance remains in the dual-ODT scheme, and is discussed in detail in Section V of Supplemental Material.

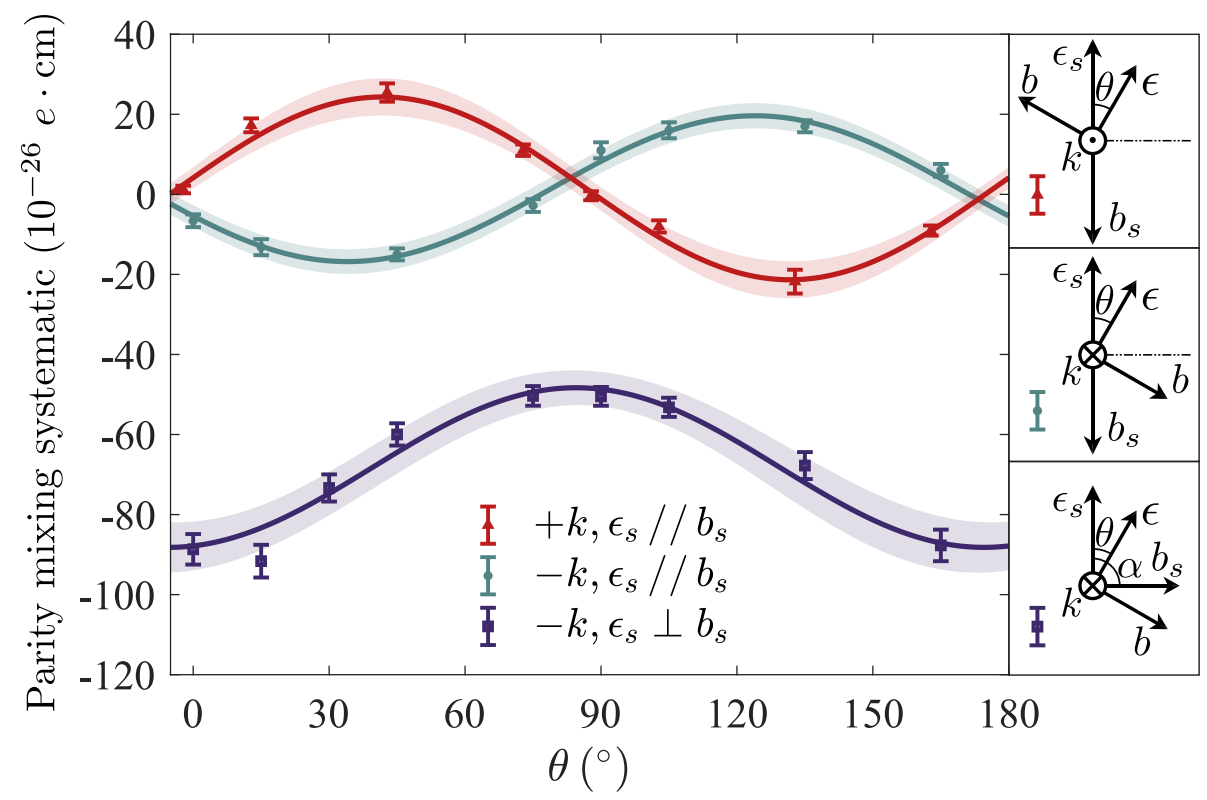

FIG. 2. Parity-mixing effect vs. ODT polarization angle $\theta$, examined under three different configurations as illustrated in the insets. Red data ($+k$, $\boldsymbol{\epsilon}_{\rm s} \parallel \boldsymbol{b}_{\rm s}$) and green data ($-k$, $\boldsymbol{\epsilon}_{\rm s} \parallel \boldsymbol{b}_{\rm s}$) show that the effect reverses sign as the ODT wave vector is inverted. Purple data ($-k$, $\boldsymbol{\epsilon}_{\rm s} \perp \boldsymbol{b}_{\rm s}$) are combined with either red or green data to extract $\mathcal{R}^{(1)}$ and $\mathcal{R}^{(2)}$ parameters. Solid lines indicate fits according to Eq. (2). The shaded areas show 95% confidence intervals.

*EDM results*—Fig. 3(a) presents the EDM results with only statistical uncertainties. The orange data points represent the EDM values averaged over ODT$_{+k,\rm H}$ and ODT$_{+k,\rm V}$, while the blue data points correspond to ODT$_{-k,\rm H}$ and ODT$_{-k,\rm V}$. The two pairs of ODTs were switched, between $+k$ and $-k$, every two days. Each data point corresponds to a run consisting of 22 measurements of the following phase combination: $[(\phi_+ - \phi_-)_{\rm V} + (\phi_+ - \phi_-)_{\rm H}]/2$, taking a total acquisition time of 74 minutes. The cumulative data collection times for the entire experiment are 1173 hours for ODT$_{+k}$ and 1175 hours for ODT$_{-k}$.

The combined EDM results with statistical uncertainties are $d_{+k} = (+4.27 \pm 0.54_{\rm stat}) \times 10^{-27}$ $e\cdot$cm with ODT$_{+k}$ and $d_{-k} = (-4.22 \pm 0.51_{\rm stat}) \times 10^{-27}$ $e\cdot$cm with ODT$_{-k}$. On one hand, the difference between the two EDM values provides information about the parity-mixing effect: $\tilde{d}_2 \cdot \sin(\alpha) = (d_{+k} - d_{-k})/2 = (4.2 \pm 0.4_{\rm stat}) \times 10^{-27}$ $e\cdot$cm ($\tilde{d}_2$ here differs from the earlier value due to a different ODT intensity). Combining the average intensity across the four ODTs, $I = (3.9 \pm 0.3) \times 10^5$ W/cm$^2$, together with the atomic structure parameters $\mathcal{R}^{(1,2)}$, the angle between the static $E$ field and the static $B$ field is determined to be $\alpha = 22(2)$ mrad. On the other hand, by averaging the $d_{+k}$ and $d_{-k}$ results, the parity-mixing systematic effect is cancelled, leading to the final statistical EDM result: $d = (d_{+k} + d_{-k})/2 = (0.3 \pm 3.7_{\rm stat}) \times 10^{-28}$ $e\cdot$cm.

The stability of the EDM measurements can be demonstrated with the Allan deviation plot (Fig. 3(b)). The dark green line represents the standard quantum limit determined by the spin projection noise: $4.3 \times 10^{-26}/\sqrt{n}$ $e\cdot$cm, where $n$ is the number of measurements. Taking into account the Johnson noise—the noise induced by Johnson currents in the copper electrodes—the combined total noise level increases to $6.7 \times 10^{-26}/\sqrt{n}$ $e \cdot \mathrm{cm}$ (light green line). Ref. [15] reported a magnetic noise of about 3 pT, which was attributed to the Seebeck effect. In the present work, after improving the field alignment to ensure that the bias B field is perpendicular to the ODT $k$ vector (see Supplemental Material IV), the magnetic-like noise is suppressed to the level of 1 pT. Therefore, we believe that the noise reported in Ref. [15] mainly originated from the ODT vector light shift. As a result, the Allan deviations of the EDM values measured in $\mathrm{ODT}_{+k}$ (orange) and $\mathrm{ODT}_{-k}$ (blue) are nearly identical, and are mainly limited by the combined effect of the quantum spin projection noise and electrode Johnson noise.

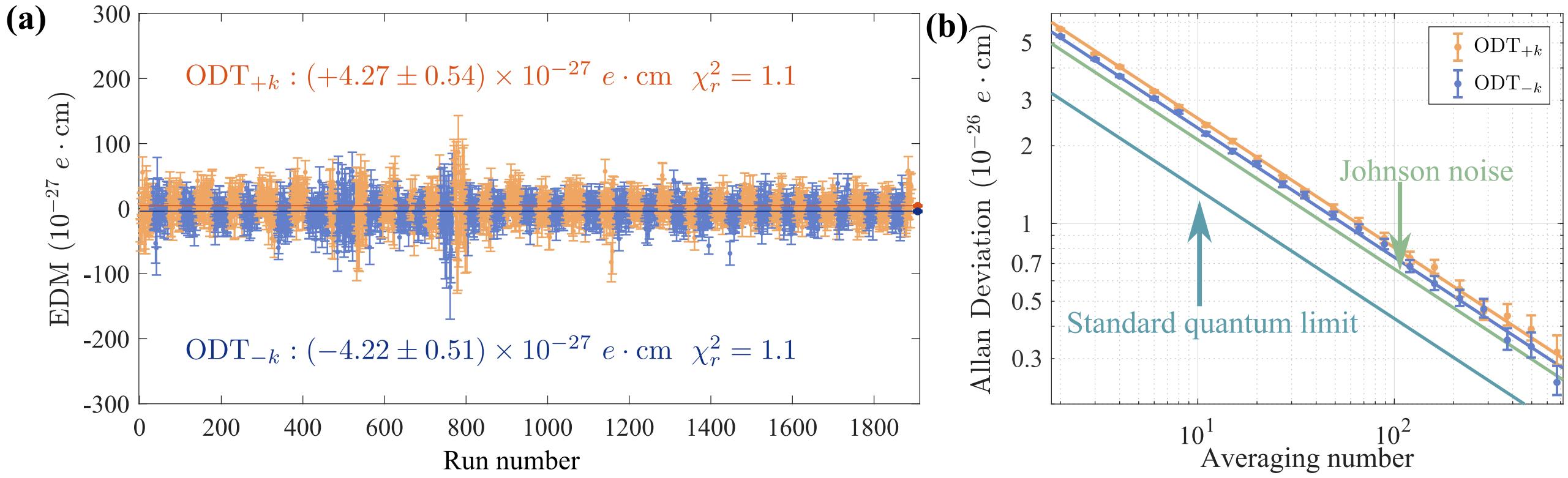

FIG. 3. (a) Measured EDM values. Orange and blue points represent results obtained in $\mathrm{ODT}_{+k}$ and $\mathrm{ODT}_{-k}$, respectively. (b) Allan deviation of the EDM results. The four lines correspond to the standard quantum limit determined by the quantum spin projection noise (dark green), the expected sensitivity when Johnson noise is included on top of the standard quantum limit (light green), and the results from $\mathrm{ODT}_{+k}$ (orange) and $\mathrm{ODT}_{-k}$ (blue).

Systematic uncertainties are summarized in Table I. The parity-mixing effect is suppressed by two orders of magnitude in two steps: first, averaging between H and V within each co-propagating pair of ODTs; second, averaging between $+k$ and $-k$ between the two counter-propagating pairs. However, there is an intensity mismatch among the four ODTs, leading to a residual systematic correction and uncertainty. The bias $B$-field correlation term is based on a hypothesized correlation between the high-voltage source and the current source for the bias $B$ field, possibly through a ground connection, potentially causing the $B$ field to vary systematically upon an $E$-field reversal. This term is limited only by the measurement precision of the electrical current for the $\cos(\theta)$ coil. A change in leakage currents between positive and negative $E$ fields is observed to be $\Delta I_{\mathrm{leak}} = 0.24$ nA. While the leakage current may actually flow through various surfaces of high-voltage insulators, we make the conservative assumption of its being a straight-line current in vacuum, generating a loop magnetic field that changes upon an $E$-field reversal. More details are provided in Supplemental Material.

Combining both statistical and systematic uncertainties, the final result for the atomic EDM of $^{171}$Yb is $d(^{171}\mathrm{Yb}) = (+0.1 \pm 3.7_{\mathrm{stat}} \pm 0.9_{\mathrm{syst}}) \times 10^{-28}$ $e \cdot \mathrm{cm}$. This corresponds to an upper limit of $|d(^{171}\mathrm{Yb})| < 7.5 \times 10^{-28}$ $e \cdot \mathrm{cm}$ (95% C.L.). Given the relation between the atomic EDM and the nuclear Schiff moment for $^{171}$Yb, $d(^{171}\mathrm{Yb}) = -2.1 \times 10^{-4}$ $S(^{171}\mathrm{Yb})/\mathrm{fm}^2$ [20], we obtain the upper limit for the nuclear Schiff moment: $|S(^{171}\mathrm{Yb})| < 3.6 \times 10^{-11}$ $e$ $\mathrm{fm}^3$.

TABLE I. Systematic effects on the measured EDM. All entries are specified in units of $10^{-28}$ $e \cdot \mathrm{cm}$.

| Contribution | Correction | Uncertainty |
|---|---|---|
| Parity-mixing effect | −0.86 | 0.35 |
| Bias $B$-field correlation | +0.67 | 0.70 |
| Leakage current | | 0.46 |
| ODT power correlation | | 0.05 |
| E-squared effect | | 0.01 |
| Total | −0.19 | 0.91 |

*Nuclear model calculation*—A recently developed multireference covariant density functional theory (MR-CDFT) [26] provides a fully relativistic, beyond-mean-field framework for describing nuclear Schiff moments [27]. In this approach, nuclear wave functions are constructed as superpositions of symmetry-restored mean-field configurations with different shapes, incorporating particle-number, parity, and angular-momentum projection. The framework captures collective quadrupole–octupole fluctuations and correlations that lie beyond static mean-field or random-phase-approximation descriptions. Using MR-CDFT, we perform the first calculation of the nuclear structure factors of the Schiff moment of $^{171}$Yb. The results indicate that it is dominated by contributions from both low-lying and high-lying excited states of opposite parity. The calculation introduces no adjustable parameters beyond those of the energy density functionals [27].

As validation, MR-CDFT is shown to accurately reproduce the energy spectra and the electromagnetic observables of $^{171}$Yb, including key spectroscopic features such as its electric quadrupole and magnetic dipole transition strengths (Supplemental Material).

The nuclear Schiff moments are sensitive to the T- and P-violating pion-nucleon couplings characterized by the isoscalar ($\bar{g}_\pi^{(0)}$), isovector ($\bar{g}_\pi^{(1)}$), and isotensor ($\bar{g}_\pi^{(2)}$) constants:

$$S = g_{\pi NN}\vec{a}\cdot\vec{g} = g_{\pi NN}\left(a_0\bar{g}_\pi^{(0)} + a_1\bar{g}_\pi^{(1)} + a_2\bar{g}_\pi^{(2)}\right), \quad (3)$$

where $g_{\pi NN} = 12.9$ is the QCD pion-nucleon coupling constant [17]. The nuclear structure factors $(a_0,\ a_1,\ a_2) = (-0.0053,\ 0.0126,\ -0.0168)\ e\cdot\mathrm{fm}^3$ for $^{171}$Yb are provided by the MR-CDFT calculation.

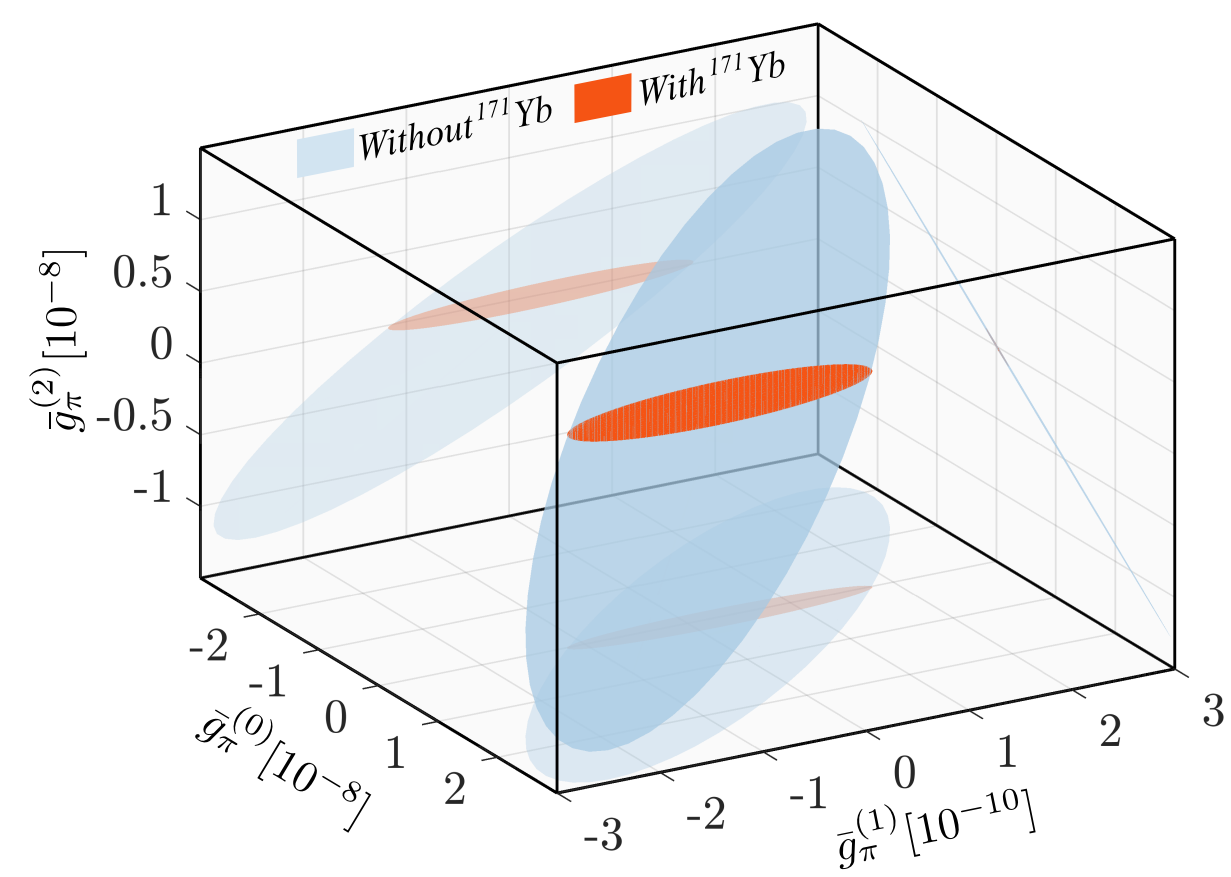

FIG. 4. Parameter space of the three CP-violating coupling constants of pion-nucleon interactions. The blue ellipsoid represents the 95% confidence limits on $\bar{g}_\pi^{(0)}$ (isoscalar), $\bar{g}_\pi^{(1)}$ (isovector), and $\bar{g}_\pi^{(2)}$ (isotensor) set by EDM measurements from $^{199}$Hg, $^{129}$Xe, $^{225}$Ra, and $^{205}$TlF [8, 11, 12, 14], without $^{171}$Yb. By including the $^{171}$Yb EDM result, the 95% confidence region is compressed to the red ellipsoid, improving the constraints on $\bar{g}_\pi^{(0)}$ and $\bar{g}_\pi^{(2)}$ by more than an order of magnitude. Note the scale for $\bar{g}_\pi^{(1)}$ is two orders of magnitude finer than that of $\bar{g}_\pi^{(0)}$ and $\bar{g}_\pi^{(2)}$, indicating that the ellipsoid is approximately in the $\bar{g}_\pi^{(0)}$–$\bar{g}_\pi^{(2)}$ plane.

Based on the experimental upper limit on $S(^{171}\mathrm{Yb})$ and the nuclear-structure calculations performed using the MR-CDFT model, both of this work and Ref. [28] (values provided in Supplemental Material, Section VII) , we derive constraints on $\bar{g}_\pi^{(0,\,1,\,2)}$. If only one of the three parameters is allowed to be nonzero, then the single-parameter limit is best set by the $^{199}$Hg result alone. On the other hand, in a global analysis of all three parameters, the $^{171}$Yb result provides a complementary constraint. According to Eq. (3), in the three-dimensional parameter space (Fig. 4), an upper limit on $S$ confines $\bar{g}_\pi^{(0,\,1,\,2)}$ values to an infinite plate normal to the vector $\vec{a}$. Combining results from three or more nuclei reduces the allowed parameter space to an ellipsoid. As shown in Fig. 4, the blue region represents the combined constraints from existing results including $^{199}$Hg, $^{129}$Xe, $^{225}$Ra, and $^{205}$TlF [8, 11, 12, 14]. By including the $^{171}$Yb result, the allowed parameter space is compressed to the red region. Interestingly, due to what appears to be an accidental relation between the nuclear structure factors of $^{199}$Hg and $^{205}$TlF, $\vec{a}_{\mathrm{Hg}}\times\vec{a}_{\mathrm{TlF}} \approx 10^{-3}(3.6, -0.03, -1.6)$, the ellipsoid lies approximately in the $\bar{g}_\pi^{(0)}$–$\bar{g}_\pi^{(2)}$ plane. The 95% confidence upper limits under different combinations of nuclear model calculations are summarized in Table II, demonstrating that inclusion of the $^{171}$Yb result significantly improves the constraints on all $\bar{g}_\pi^{(0,\,1,\,2)}$.

TABLE II. Constraints on $\bar{g}_\pi^{(0,\,1,\,2)}$ with and without the $^{171}$Yb atomic EDM result. The last column labels the nuclear models used to calculate $(a_0,\ a_1,\ a_2)$. Atomic and molecular-structure parameters for $^{171}$Yb, $^{199}$Hg, $^{129}$Xe, $^{225}$Ra, and $^{205}$TlF are taken from Refs. [20, 29–32], and EDM results are from Refs. [8, 11, 12, 14].

| | Constraints [$10^{-10}$] | | | Nuclear models |
|---|---|---|---|---|
| | $\bar{g}_\pi^{(0)}$ | $\bar{g}_\pi^{(1)}$ | $\bar{g}_\pi^{(2)}$ | |
| Without $^{171}$Yb | 193 | 2.0 | 86 | MR-CDFT [28] |
| With $^{171}$Yb | 13 | 1.2 | 5.9 | |
| Without $^{171}$Yb | 159 | 3.9 | 55 | Multiple models[a] |
| With $^{171}$Yb | 37 | 0.9 | 13 | |

[a] $(a_0,\ a_1,\ a_2)$ for $^{199}$Hg and $^{129}$Xe are obtained from pair-truncated shell–model calculations [33, 34]. The values for $^{225}$Ra were calculated using Skyrme DFT [35]; those for $^{205}$TlF are taken from Ref. [17]; and those for $^{171}$Yb from this work.

*Discussion and outlook*—To further improve the search sensitivity on $^{171}$Yb EDM towards $10^{-29}\,e\cdot$cm, the Johnson noise can be reduced by exploring alternative materials for the electrodes, for example, niobium with a lower conductivity [36] or glass substrates coated with a thin conductive layer. It is also possible to use a $^{173}$Yb–$^{171}$Yb cold-atom co-magnetometer [37] to suppress the magnetic-field noise. The parity-mixing effects can be further suppressed by replacing running-wave ODTs with optical lattices in optical cavities, which would ensure better intensity matching between $+k$ and $-k$ waves.

Given the similarity in atomic structure between Yb and Ra, the methods developed here can be directly applied to $^{225}$Ra, a radioactive isotope with a 15-day half-life. Owing to its octupole-deformed nucleus, $^{225}$Ra exhibits exceptional sensitivity to hadronic CP-violating parameters [35]—about three orders of magnitude higher than that of $^{171}$Yb or $^{199}$Hg. Measuring the EDM of $^{225}$Ra at the precision level of $10^{-28}\,e\cdot$cm would then realize the most sensitive search for BSM physics in the nuclear Schiff-moment category.

*Acknowledgments*—We thank D. Sheng, W. Jiang, W.-T. Luo, and J.-L. Zhang for fruitful discussions, and acknowledge the contributions of T. A. Zheng and Y. A. Yang to the early-stage development. This work is supported by the National Natural Science Foundation of China (NSFC) through grant no. 12174371, the Quantum Science and Technology-National Science and Technology Major Project through grant no. 2021ZD0303101, the Strategic Priority Research Program of the Chinese Academy of Sciences through grant no. XDB21010200, and the Major Frontier Research Project of the University of Science and Technology of China through grant no. LS9990000002. J.-M. Yao is supported by NSFC through grant no. 12375119.

*Data availability*—The data that support the findings of this study are available from the corresponding author upon reasonable request, and have also been deposited in [38].

# Supplemental Material for "An improved limit on $^{171}$Yb electric dipole moment"

S.-Z. Wang[1],* S.-B. Wang[1],* E. F. Zhou[2,3], Z.-J. Tao[1], J.-M. Yao[2,3],† T. Xia[4],‡ and Z.-T. Lu[1,4]§

[1] *CAS Center for Excellence in Quantum Information and Quantum Physics, School of Physical Sciences, University of Science and Technology of China, Hefei, Anhui, 230026, China.*

[2] *School of Physics and Astronomy, Sun Yat-sen University, Zhuhai, Guangdong, 519082, China.*

[3] *Guangdong Provincial Key Laboratory of Quantum Metrology and Sensing, Sun Yat-sen University, Zhuhai, Guangdong, 519082, China.* and

[4] *Hefei National Laboratory, University of Science and Technology of China, Hefei 230088, China*

## I. ODT INTENSITY CALIBRATION

Atoms confined in an ODT experience an AC Stark shift whose magnitude depends on the ODT intensity $I$ and the differential dynamic polarizability $\Delta\alpha(\lambda)$ between the ground and excited states. The shift is given by $h\Delta\nu_{\mathrm{AC}} = -\Delta\alpha(\lambda)I/(2c\varepsilon_0)$. Near the magic wavelength $\lambda_0$, $\Delta\alpha(\lambda)$ can be approximated by $\Delta\alpha(\lambda) = c_0 + c_1(\lambda - \lambda_0) + c_2(\lambda - \lambda_0)^2$, where $c_i$ $(i = 0, 1, 2)$ are atomic structure parameters. By combining the measured AC Stark shifts at different ODT wavelengths with the calculated $c_1$ value, the ODT intensity averaged over the atomic ensemble can be determined.

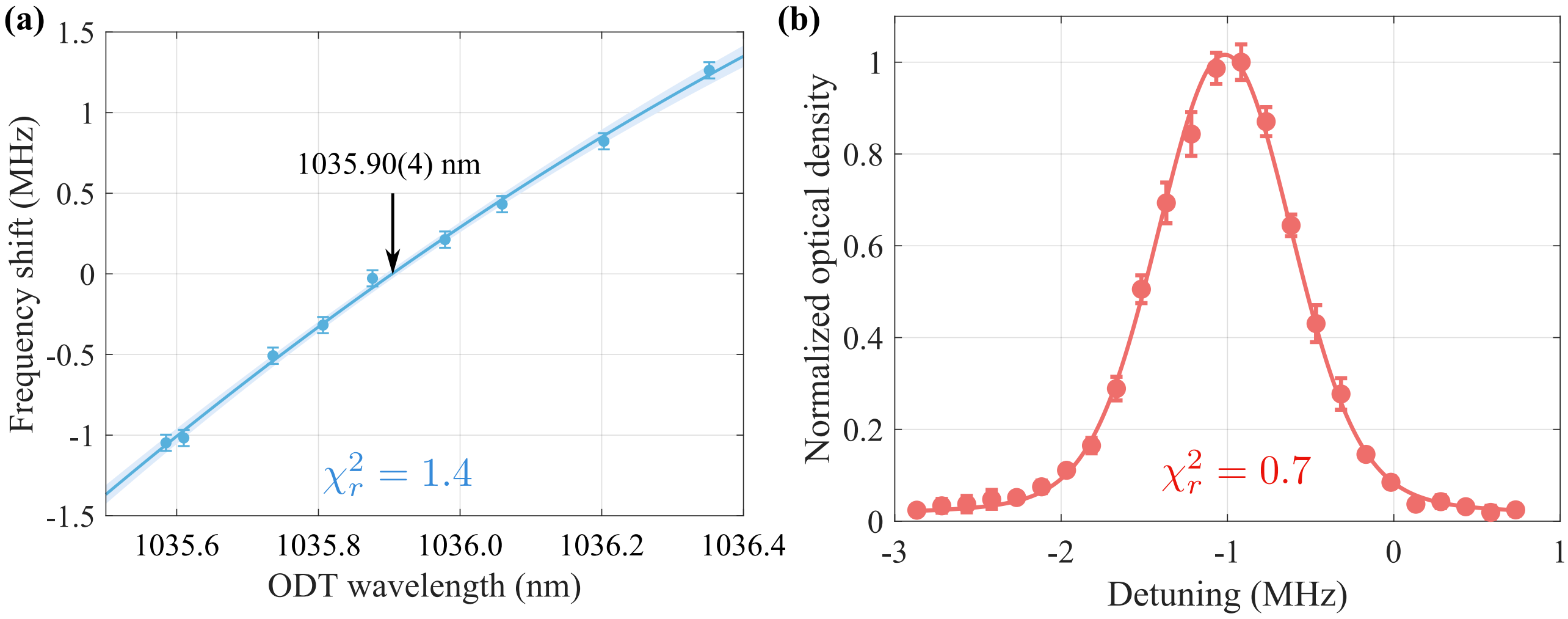

Figure S1. (a) AC Stark shift vs. ODT wavelength. Each data point is obtained from a spectral scan (see (b) as an example) taken with the ODT wavelength set at the data-point value. A fit to the data yields a slope of 3.09(9) MHz/nm at the magic wavelength 1035.90(4) nm. The uncertainty in the magic wavelength is determined by the wavemeter accuracy. (b) Absorption spectrum of the $6s^2\,{}^1S_0 \rightarrow 6s6p\,{}^3P_1$ transition at 556 nm when the ODT wavelength is fixed at $\lambda$= 1035.61 nm. The zero point of the horizontal axis corresponds to the line center measured with the ODT turned off. The absorption spectrum is fitted with a Voigt profile, where the Lorentzian component is fixed at 0.18 MHz (natural linewidth), and the Gaussian width is treated as a free parameter. The fit yields a line center of $-1.02(5)$ MHz and a linewidth of 0.88 MHz.

When studying the parity-mixing effect, the ODT intensity is set to a relatively high level to enhance the signal. We measure the AC Stark shift of the $6s^2\,{}^1S_0\,|1/2, 1/2\rangle \rightarrow 6s6p\,{}^3P_1\,|3/2, 3/2\rangle$ transition (quantization axis along z, the ODT propagation direction) in $^{171}$Yb at different ODT wavelengths. The results are shown in Fig. S1. A fit to the data yields a linear dependence at the magic wavelength, $d\Delta\nu_{\mathrm{AC}}/d\lambda|_{\lambda=\lambda_0} = -c_1 I/(2c\varepsilon_0) = 3.09(9)$ MHz/nm. Given the theoretical value $c_1 = -1.13(8)\times 10^{-12}$ MHz/(V/m)$^2$/nm [1], we obtain the average ODT intensity over the atomic ensemble to be $I = (1.45 \pm 0.11)\times 10^6$ W/cm$^2$.

* These authors contributed equally to this work.
† Corresponding author: yaojm8@sysu.edu.cn
‡ Corresponding author: txia@hfnl.cn
§ Corresponding author: ztlu@ustc.edu.cn

In the EDM measurement, the ODT intensities are reduced in order to suppress the parity-mixing effect. Using the same calibration procedure, we determine the intensities of the four ODTs; their average value is $(3.9 \pm 0.3)\times 10^5$ W/cm$^2$.

## II. ELECTRIC FIELD CALIBRATION

The static electric field $E_s$ is generated across a pair of parallel-plate electrodes of 10 mm diameter and 0.76(2) mm spacing. The voltage is monitored through a voltage divider with a 0.01% precision. We define $\kappa$ as the conversion factor between the electric field and the applied voltage ($E_s = \kappa V$). Although $\kappa$ can be directly calculated from the measured electrode spacing, we employ a more precise calibration method based on DC Stark shift measurements.

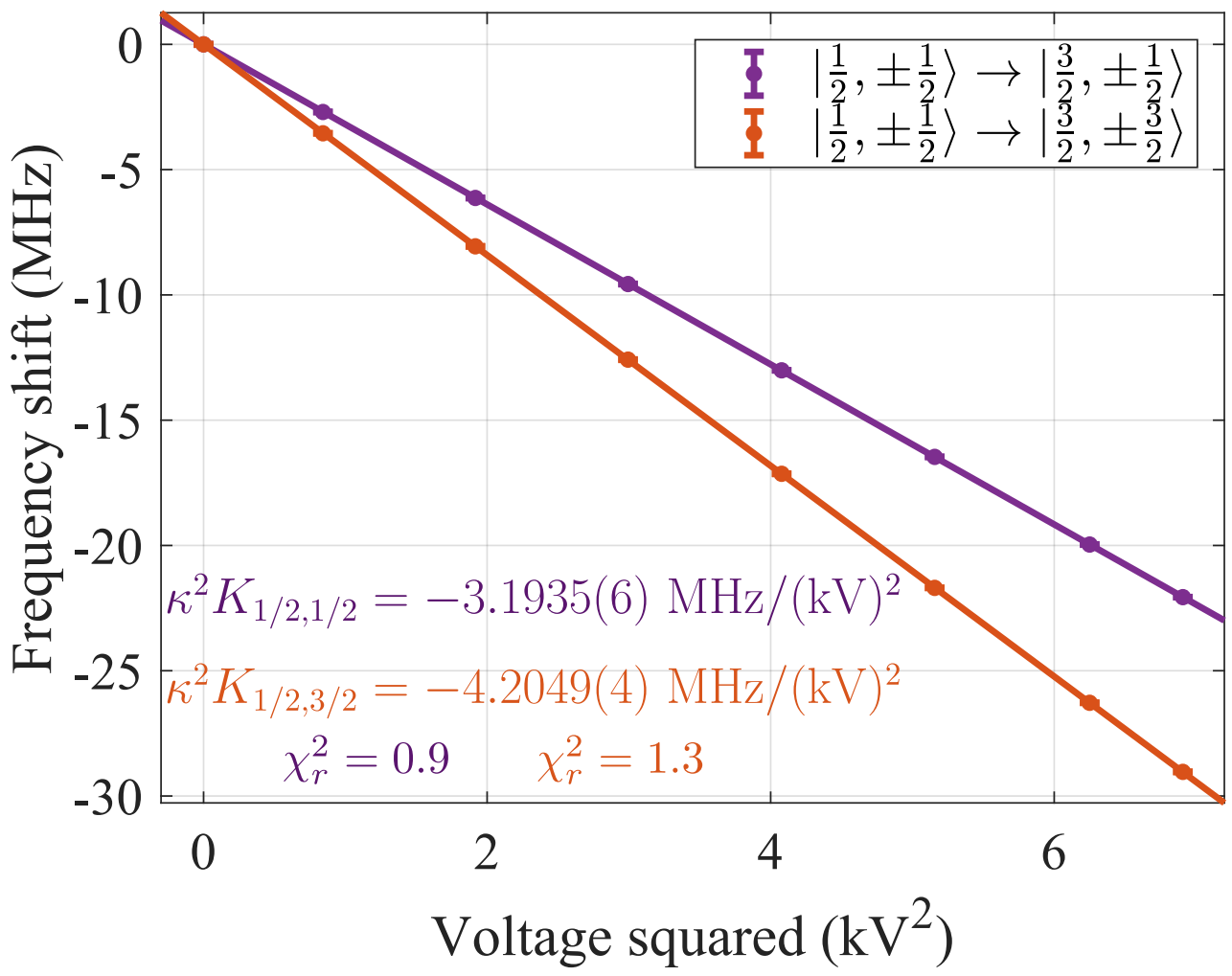

Figure S2. DC Stark shifts of the $6s^2\ ^1S_0\ F=1/2 \to 6s6p\ ^3P_1\ F=3/2$ transition in $^{171}$Yb. The purple line corresponds to the $|1/2,\pm1/2\rangle \to |3/2,\pm1/2\rangle$ transition, and the orange line to $|1/2,\pm1/2\rangle \to |3/2,\pm3/2\rangle$.

The DC Stark shift of $6s^2\,^1S_0 \to 6s6p\,^3P_1$ is given by $h\Delta\nu = K_{m,m'}E^2 = \kappa^2 K_{m,m'}V^2$, where $K_{m,m'}$ is the DC Stark-shift coefficient and $m$ ($m'$) denotes the magnetic sublevel of the ground (excited) state. Following Ref. [2], we adopt the reference value $K_{0,0} = -1.5419(48)\,$MHz/(kV/mm)$^2$ of $^{176}$Yb [3]. In this work, we measure the DC Stark shifts of $^{171}$Yb, for which the coefficients obey the relation $2K_{0,0} = 3K_{1/2,1/2} - K_{1/2,3/2}$, where $K_{1/2,1/2}$ corresponds to the $|1/2,\pm1/2\rangle \to |3/2,\pm1/2\rangle$ transition and $K_{1/2,3/2}$ to $|1/2,\pm1/2\rangle \to |3/2,\pm3/2\rangle$.

We measure the DC Stark shifts at various voltages, obtaining slopes of $\kappa^2 K_{1/2,1/2} = -3.1935(6)\,$MHz/kV$^2$ and $\kappa^2 K_{1/2,3/2} = -4.2049(4)\,$MHz/kV$^2$ (Fig. S2). By combining these measured slopes with the reference value $K_{0,0}$, we obtain the electric-field conversion factor $\kappa = 1.3203(21)\,$(kV/mm)/kV.

## III. FIELD ALIGNMENT

In order to suppress systematic effects, we precisely tune the relative orientations of three vectors: the magnetic field direction ($\boldsymbol{b_s}$), the electric field direction ($\boldsymbol{\epsilon_s}$), and the ODT propagation direction ($\boldsymbol{e_k}$). We first align both the magnetic and electric fields to be perpendicular to the ODT propagation direction ($\boldsymbol{b_s} \perp \boldsymbol{e_k}$, $\boldsymbol{\epsilon_s} \perp \boldsymbol{e_k}$), ensuring that the static magnetic field, static electric field, optical electric field, and optical magnetic field all lie within the same plane. We then finely adjust the static magnetic field so that it becomes parallel to the static electric field ($\boldsymbol{b_s} \parallel \boldsymbol{\epsilon_s}$).

### A. $b_s \perp e_k$

The ODT beam propagates along the $z$ direction ($\boldsymbol{e_k} = \boldsymbol{e_z}$), as do the polarization beam, the dressing beam and the probe beam. The angular deviations among these four beams remain below 1 mrad.

Let us first assume the ideal case of $\boldsymbol{b_s} \perp \boldsymbol{e_k}$. In QND measurements, the first measurement yields the population $\rho_{11}^{(1)}$ of the $|m = +1/2\rangle$ state (quantization along $z$-direction). Following half a Larmor period, a second population measurement yields $\rho_{11}^{(2)}$, which is equal to $\rho_{22}^{(1)}$, the population of the $|m = -1/2\rangle$ state during the first measurement. Thus, the two populations $\rho_{11}$ and $\rho_{22}$ are measured alternately every half-period. Under the assumed perfect perpendicularity, the populations are preserved under repeated QND measurements.

If $\boldsymbol{b_s}$ is not perpendicular to $\boldsymbol{e_k}$, atoms initially in $|m = -1/2\rangle$ do not evolve into $|m = +1/2\rangle$ after half a period. Instead, they evolve into a superposition of $|m = +1/2\rangle$ and $|m = -1/2\rangle$, and repeated measurement cycles cause $\rho_{11}$ and $\rho_{22}$ to converge—a depolarization effect. The depolarization rate provides an error signal for optimizing the perpendicularity between $\boldsymbol{b_s}$ and $\boldsymbol{e_k}$.

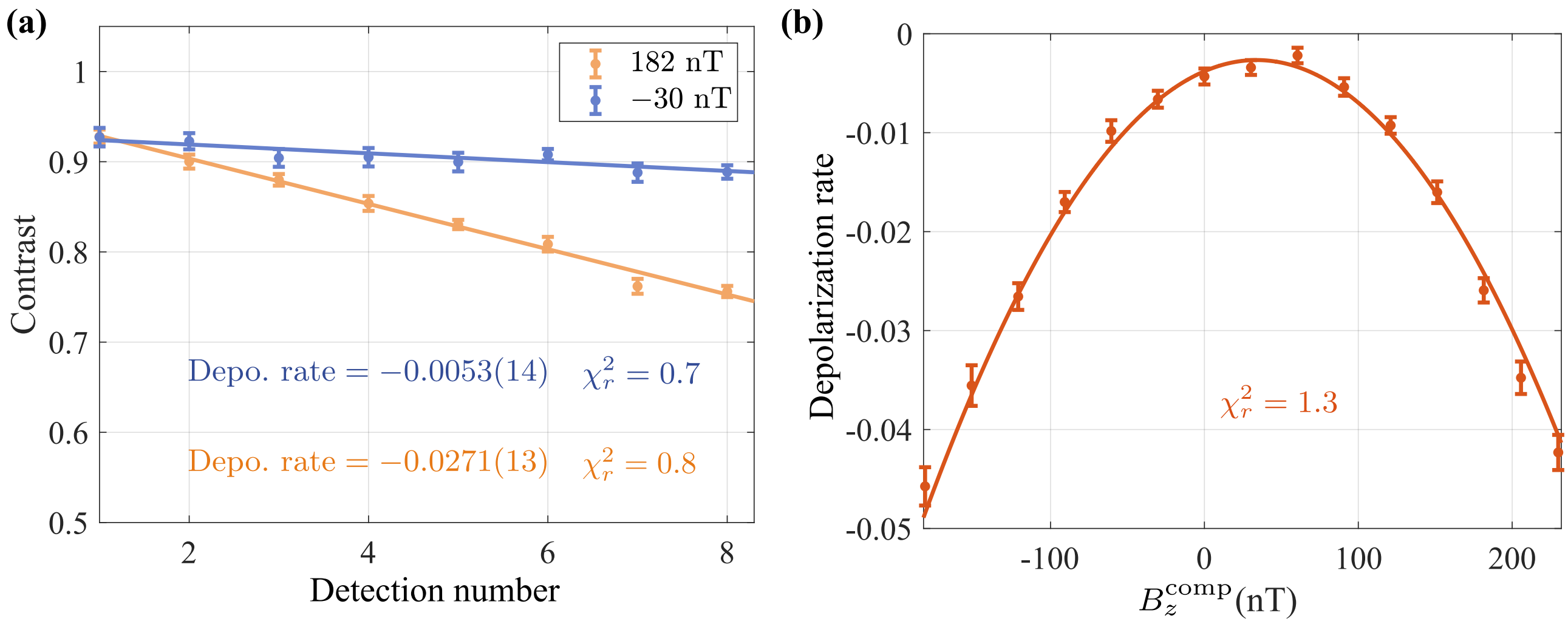

Figure S3. (a) Precession contrast decays as spin-state detections are performed. Linear fits give the depolarization rates. The orange line corresponds to a compensating magnetic field $B_z^{\mathrm{comp}} = 182$ nT, and the blue line corresponds to $B_z^{\mathrm{comp}} = 30$ nT. (b) Depolarization rate vs. $B_z^{\mathrm{comp}}$. A quadratic fit yields an extremum at +34(2) nT, corresponding to the condition where the total magnetic field is perpendicular to the ODT propagation direction.

We define the angle between $\boldsymbol{b_s}$ and $\boldsymbol{e_k}$ as $\pi/2+\beta$, where $\beta$ is the small deviation angle. In addition to the main magnetic field $B \approx 2000\,\mathrm{nT}$, we apply a compensating field $B_z^{\mathrm{comp}}$ along the $z$-direction to tune $\beta$ toward zero. In the experiment, a pair of population measurements (spin detection) yields the spin projection $S_z$ and reduces $S_z$ by a factor of $\Delta S_z/S_z = -4\sin^2(\beta) = -4(B_z^{\mathrm{comp}} - B_z)^2/B^2$ due to the depolarization effect (Fig. S3(a))(see Section IV for details). As shown in Fig. S3(b), the depolarization rate depends quadratically on $B_z^{\mathrm{comp}}$. The extremum of the quadratic fit gives $B_z^{\mathrm{comp}} = +34(2)\,\mathrm{nT}$, at which the magnetic field is perpendicular to the ODT propagation direction. Following this procedure, the deviation from perfect perpendicularity is tuned to below 1 mrad.

### B. $\epsilon_s \perp e_k$

The ODT passes through the center of the gap between the electrodes. Because the electrode diameter is 10 mm while the spacing is only 0.76 mm, the angle between $\boldsymbol{e_k}$ and the electrode surface is constrained. By illuminating the trapped atoms with a resonant 556 nm beam along the $x$-direction, the images of an elongated atom cloud and the electrode edges are analyzed, and the angle between $\boldsymbol{e_k}$ and the electrode surface is determined to be $(0.1 \pm 0.1)$ mrad.

### C. $b_s \parallel \epsilon_s$

Following the above tuning procedure, both the magnetic and electric fields lie in the $x$-$y$ plane. As described in the main text, the angle $\alpha$ between the two fields is related to the parity-mixing signal $\tilde{d}_2 \sin\alpha = (d_{+k} - d_{-k})/2$. This allows us to align the magnetic field to minimize $d_{+k} - d_{-k}$. During the EDM data-taking period, we obtain $(d_{+k} - d_{-k})/2 = (4.2 \pm 0.4) \times 10^{-27}$ $e \cdot$cm. Using the calibrated ODT intensity and the atomic coefficients $\mathcal{R}^{(1,2)}$ determined from the parity-mixing measurements, we find $\tilde{d}_2 = (1.9 \pm 0.1) \times 10^{-25}$ $e \cdot$cm, which leads to a residual field misalignment angle $\alpha = 22(2)$ mrad.

## IV. MAGNETIC FIELD ALIGNMENT VIA QND SPIN MEASUREMENTS

The EDM measurement in this work requires the magnetic field to be perpendicular to the propagation direction of the ODT. To precisely align the magnetic field, we employ the spin-selective and spin-preserving QND measurement technique [4, 5], and apply the observed spin depolarization rate as a feedback signal. Both the probe and the dressing laser beams used in the QND measurement are colinear with the ODT beam.

### A. Bloch-vector interpretation

As shown in Fig. S4, both the probe and dressing beams propagate along the $z$-direction. The magnetic field ($B$) is slightly misaligned from the $y$-direction by an angle $\beta$ ($\beta \ll 1$). Initially, the Bloch vector ($S$) points along $+z$ with a magnitude $S_z = (N_+ - N_-)/(N_+ + N_-)$, where $N_\pm$ represent the populations of spins pointing toward $\pm z$. Under the magnetic field, $S$ precesses around $B$. After a half precession period, the angle between $S$ and $-z$ becomes $2\beta$. At this point, a QND measurement of $N_+$ projects $S$ back onto the $z$-axis, reducing its magnitude to $\cos(2\beta) \cdot S_z$. Similarly, after the subsequent half-period of precession, a QND measurement of $N_-$ projects $S$ onto the $z$-axis again, and further reduces its magnitude to $\cos^2(2\beta) \cdot S_z$. Therefore, the relative change induced by two consecutive projection measurements can be written as

$$\frac{\Delta S_z}{S_z} = -\sin^2(2\beta) \approx -4\sin^2(\beta). \tag{1}$$

In the QND measurement, as the spin-projection operations are repeated, the depolarization effect described above leads to a continuous reduction of the Bloch-vector magnitude.

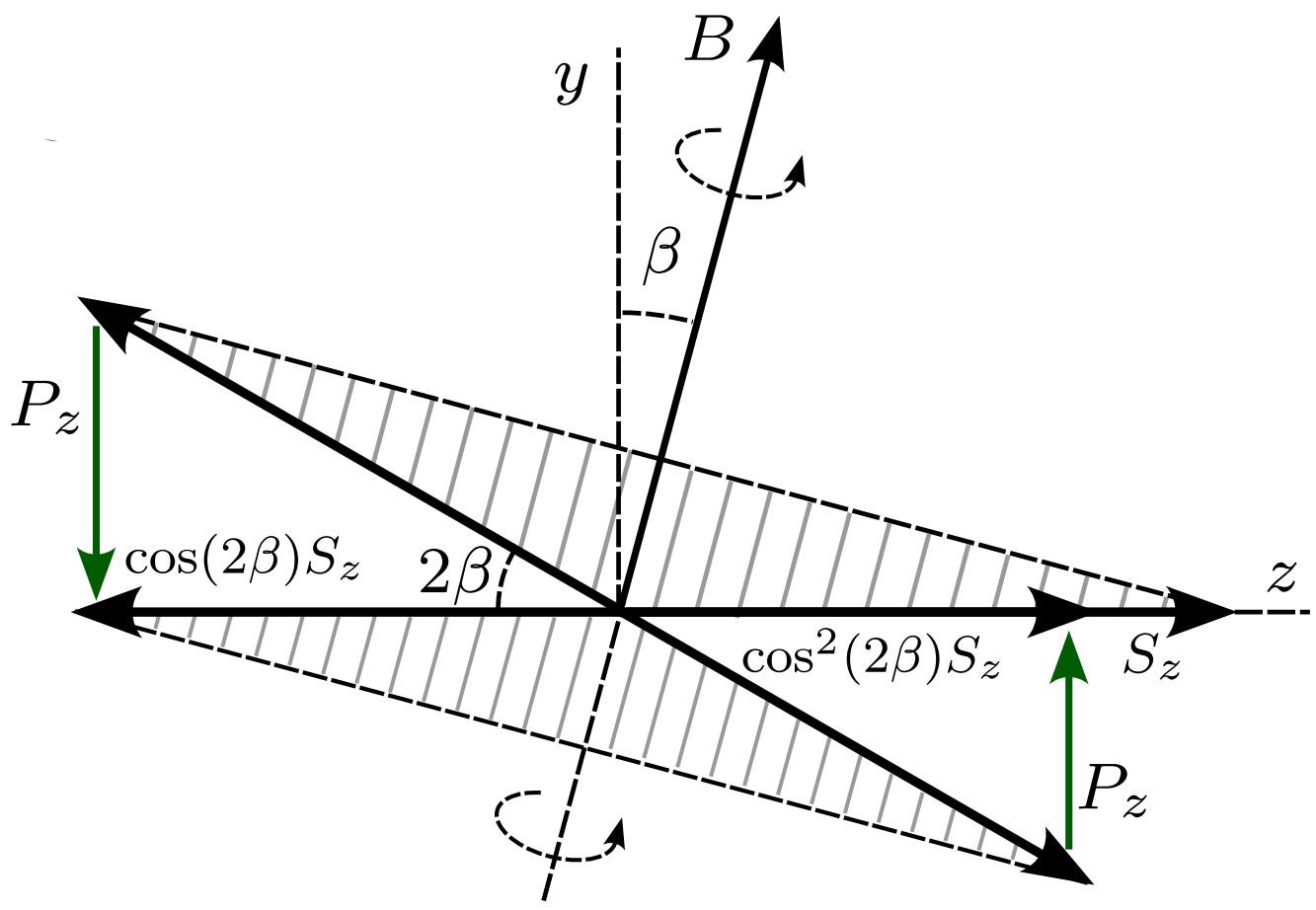

Figure S4. Spin evolution in the QND measurement.

### B. Density-matrix description

The QND measurement process described above can also be formulated in terms of density-matrix evolution. Defining the $z$-axis in Fig. S4 as the quantization axis, the Hamiltonian describing the interaction between the spin and the magnetic field is $H = \hbar\omega/2\,(\cos\beta\,\sigma_y + \sin\beta\,\sigma_z)$, where $\omega$ is the Larmor precession frequency. Initially, the density matrix is $\rho^{(0)} = \mathrm{diag}\Big(\rho_{11}^{(0)},\, 1-\rho_{11}^{(0)}\Big)$, where $\rho_{11}^{(0)}$ denotes the population of the $|m_F = +1/2\rangle$ state. The density matrix then evolves in the magnetic field for half a precession period,

$$\rho^{(1)} = U_\pi\,\rho^{(0)}\,U_\pi^\dagger = \begin{pmatrix} -i\sin\beta & -\cos\beta \\ \cos\beta & i\sin\beta \end{pmatrix} \begin{pmatrix} \rho_{11}^{(0)} & 0 \\ 0 & 1-\rho_{11}^{(0)} \end{pmatrix} \begin{pmatrix} i\sin\beta & \cos\beta \\ -\cos\beta & -i\sin\beta \end{pmatrix} = \begin{pmatrix} \rho_{11}^{(1)} & \rho_{12}^{(1)} \\ \rho_{21}^{(1)} & \rho_{22}^{(1)} \end{pmatrix}, \tag{2}$$

where $U_\pi$ is the evolution operator corresponding to a $\pi$ rotation. After the QND measurement, the off-diagonal elements of the density matrix collapse to zero, and the population is found to be $\rho_{11}^{(1)} = -\cos(2\beta)\,\rho_{11}^{(0)} + \cos^2\beta$. Similarly, after the subsequent half-period of precession followed by another projective measurement, the population becomes $\rho_{11}^{(2)} = -\cos(2\beta)\,\rho_{11}^{(1)} + \cos^2\beta$.

In the experiment, each pair of successive projections is used to extract one value of the spin projection $S_z$. The first measured spin projection can be written as

$$S_z^{(1)} = \frac{\rho_{11}^{(1)} - \rho_{11}^{(2)}}{\rho_{11}^{(1)} + \rho_{11}^{(2)}} \approx \rho_{11}^{(1)} - \rho_{11}^{(2)} = \cos^2\beta\,\Big(2\rho_{11}^{(1)} - 1\Big). \tag{3}$$

Similarly, the $i$-th spin projection is given by

$$S_z^{(i)} \approx \rho_{11}^{(2i-1)} - \rho_{11}^{(2i)} = \cos^{2i-2}(2\beta)\,\cos^2\beta\,\Big(2\rho_{11}^{(1)} - 1\Big) = \cos^{2i-2}(2\beta)\,S_z^{(1)}. \tag{4}$$

The relative change between two consecutive spin measurements is therefore

$$\frac{\Delta S_z}{S_z^{(i-1)}} = \frac{S_z^{(i)} - S_z^{(i-1)}}{S_z^{(i-1)}} = -\sin^2(2\beta). \tag{5}$$

This result shows that the spin projection $S_z$ decays as the number of projective measurements increases, with a decay rate of $\sin^2(2\beta)$, in agreement with the conclusion obtained from the Bloch-vector interpretation.

## V. SYSTEMATIC ERROR ANALYSIS

### A. Residual parity-mixing effect

We first define the following symbols: $d(^{171}\mathrm{Yb})$ denotes the EDM of $^{171}\mathrm{Yb}$; $d$ is the raw value of the EDM measurement including systematic effects; while $\tilde{d}$ represents the false EDM arising from parity mixing effects, and is distinguished from $d$ by adding a tilde. These quantities satisfy the following relation: $d = d(^{171}\mathrm{Yb}) + \tilde{d}$. Furthermore, since the angle between H and V polarizations deviates from $\pi/2$ by less than 0.1 mrad, its contribution to the parity-mixing systematic uncertainty is negligible.

Following Eq. (2) in the main text, the parity-mixing effect in the four ODTs can be expressed as

$$\begin{aligned} \tilde{d}_{+k,\mathrm{V}} &= \tilde{d}_1\sin(2\theta_{+k} - \alpha) + \tilde{d}_2\sin(\alpha), \\ \tilde{d}_{+k,\mathrm{H}} &= \tilde{d}_1\sin(2(\theta_{+k} + \pi/2) - \alpha) + \tilde{d}_2\sin(\alpha) = -\tilde{d}_1\sin(2\theta_{+k} - \alpha) + \tilde{d}_2\sin(\alpha), \\ \tilde{d}_{-k,\mathrm{V}} &= -\tilde{d}_1\sin(2\theta_{-k} - \alpha) - \tilde{d}_2\sin(\alpha), \\ \tilde{d}_{-k,\mathrm{H}} &= \tilde{d}_1\sin(2\theta_{-k} - \alpha) - \tilde{d}_2\sin(\alpha), \end{aligned} \tag{6}$$

where the parity-mixing effects for $\mathrm{ODT}_{+k,\mathrm{V}}$, $\mathrm{ODT}_{+k,\mathrm{H}}$, $\mathrm{ODT}_{-k,\mathrm{V}}$, and $\mathrm{ODT}_{-k,\mathrm{H}}$ are denoted as $\tilde{d}_{+k,\mathrm{V}}$, $\tilde{d}_{+k,\mathrm{H}}$, $\tilde{d}_{-k,\mathrm{V}}$, and $\tilde{d}_{-k,\mathrm{H}}$, respectively. In addition, $\theta_{+k}$ is the angle between the polarization of $\mathrm{ODT}_{+k,\mathrm{V}}$ and the static $E$ field, $\theta_{-k}$ is the corresponding angle for $\mathrm{ODT}_{-k,\mathrm{V}}$, and $\alpha$ is the angle between the static $E$ and $B$ fields.

Let's first assume the ideal case that all ODTs have equal intensities. As can be seen from Eq. (6), for the two ODTs along the $+z$ direction, their average is $(\tilde{d}_{+k,V} + \tilde{d}_{+k,H})/2 = \tilde{d}_2 \sin(\alpha)$. This shows that averaging over H- and V-polarized ODTs cancels the $\theta$-dependent term. Similarly, for the two ODTs along the $-z$ direction, $(\tilde{d}_{-k,V} + \tilde{d}_{-k,H})/2 = -\tilde{d}_2 \sin(\alpha)$. Therefore, averaging over all four configurations yields: $(\tilde{d}_{+k,V} + \tilde{d}_{+k,H} + \tilde{d}_{-k,V} + \tilde{d}_{-k,H})/4 = 0$.

As shown in Fig. 3(a) of the main text, the measurements obtained with ODT$_{+k}$ or ODT$_{-k}$ individually exhibit nonzero central values, whereas this offset is largely canceled when the results from all four ODT configurations are averaged. In the actual experiment, a residual effect remains due to the uncertainties in the ODT intensity calibration and in the angles $\theta_{+k}$, $\theta_{-k}$, and $\alpha$.

The terms containing $\theta_{+k}$, $\theta_{-k}$ and $\alpha$ can be extracted from combinations of $d_{+k,\mathrm{V}}$, $d_{+k,\mathrm{H}}$, $d_{-k,\mathrm{V}}$ and $d_{-k,\mathrm{H}}$, which denote the EDM values measured in ODT$_{+k,\mathrm{V}}$, ODT$_{+k,\mathrm{H}}$, ODT$_{-k,\mathrm{V}}$, and ODT$_{-k,\mathrm{H}}$, respectively. Their relations are as follows:

$$\begin{aligned}
\tilde{d}_1 \sin(2\theta_{+k} - \alpha) &= \frac{\tilde{d}_{+k,\mathrm{V}} - \tilde{d}_{+k,\mathrm{H}}}{2} = \frac{d_{+k,\mathrm{V}} - d_{+k,\mathrm{H}}}{2}, \\
\tilde{d}_1 \sin(2\theta_{-k} - \alpha) &= -\frac{\tilde{d}_{-k,\mathrm{V}} - \tilde{d}_{-k,\mathrm{H}}}{2} = -\frac{d_{-k,\mathrm{V}} - d_{-k,\mathrm{H}}}{2}, \\
\tilde{d}_2 \sin(\alpha) &= \frac{(\tilde{d}_{+k,\mathrm{V}} + \tilde{d}_{+k,\mathrm{H}}) - (\tilde{d}_{-k,\mathrm{V}} + \tilde{d}_{-k,\mathrm{H}})}{4} = \frac{(d_{+k,\mathrm{V}} + d_{+k,\mathrm{H}}) - (d_{-k,\mathrm{V}} + d_{-k,\mathrm{H}})}{4}.
\end{aligned} \tag{7}$$

To suppress the parity-mixing systematic error, we average the measurements across the four distinct ODT configurations. The residual parity-mixing systematic error caused by the intensity imbalance of the four ODTs is given by

$$\tilde{d} = \frac{\tilde{d}_{+k,\mathrm{V}} + f_{+k,\mathrm{H}}\tilde{d}_{+k,\mathrm{H}} + f_{-k,\mathrm{V}}\tilde{d}_{-k,\mathrm{V}} + f_{-k,\mathrm{H}}\tilde{d}_{-k,\mathrm{H}}}{4}, \tag{8}$$

where $f_{+k,\mathrm{H}}$, $f_{-k,\mathrm{V}}$, and $f_{-k,\mathrm{H}}$ are the light intensity ratios of ODT$_{+k,\mathrm{H}}$, ODT$_{-k,\mathrm{V}}$, and ODT$_{-k,\mathrm{H}}$ relative to ODT$_{+k,\mathrm{V}}$, respectively, with their values listed in Table I.

Substituting Eq. (6) and Eq. (7) into Eq. (8), we can rewrite the residual parity-mixing systematic error as

$$\begin{aligned}
\tilde{d} = &\frac{(1 - f_{+k,\mathrm{H}})(d_{+k,\mathrm{V}} - d_{+k,\mathrm{H}})}{8} + \frac{(f_{-k,\mathrm{V}} - f_{-k,\mathrm{H}})(d_{-k,\mathrm{V}} - d_{-k,\mathrm{H}})}{8} \\
&+ \frac{(1 + f_{+k,\mathrm{H}} - f_{-k,\mathrm{V}} - f_{-k,\mathrm{H}})[(d_{+k,\mathrm{V}} + d_{+k,\mathrm{H}}) - (d_{-k,\mathrm{V}} + d_{-k,\mathrm{H}})]}{16}.
\end{aligned} \tag{9}$$

In the experiment, $d_{+k,\mathrm{V}}$ and $d_{+k,\mathrm{H}}$ are measured simultaneously and their uncertainties are correlated. Therefore, their combination values are used in data analysis (Table II). The same reasoning applies to $d_{-k,\mathrm{V}}$ and $d_{-k,\mathrm{H}}$.

Table I. The light intensity ratios among the four ODTs are defined as follows: $f_{+k,\mathrm{H}}$ corresponds to the ratio of ODT$_{+k,\mathrm{H}}$ to ODT$_{+k,\mathrm{V}}$, $f_{-k,\mathrm{V}}$ to that of ODT$_{-k,\mathrm{V}}$ to ODT$_{+k,\mathrm{V}}$, and $f_{-k,\mathrm{H}}$ to that of ODT$_{-k,\mathrm{H}}$ to ODT$_{+k,\mathrm{V}}$.

| $f_{+k,\mathrm{H}}$ | $f_{-k,\mathrm{V}}$ | $f_{-k,\mathrm{H}}$ |
|---|---|---|
| 1.074(4) | 0.985(14) | 1.055(21) |

Table II. Combinations of EDM measurements from the four ODTs. Values are in units of $10^{-27}$ $e \cdot$ cm.

| $(d_{+k,\mathrm{V}} - d_{+k,\mathrm{H}})$ | $(d_{+k,\mathrm{V}} + d_{+k,\mathrm{H}})$ | $(d_{-k,\mathrm{V}} - d_{-k,\mathrm{H}})$ | $(d_{-k,\mathrm{V}} + d_{-k,\mathrm{H}})$ |
|---|---|---|---|
| $-9.46 \pm 0.84$ | $8.54 \pm 1.08$ | $4.24 \pm 1.44$ | $-8.44 \pm 1.02$ |

Based on Eq. (9) and the data in Tables I and II, we obtain the parity-mixing systematic error as

$$\tilde{d} = (0.86 \pm 0.35) \times 10^{-28}\, e \cdot \mathrm{cm}.$$

### B. cos($\theta$) current

If the high-voltage supply is coupled somehow to the current supply for the cos($\theta$) coil, for example through a ground loop, the bias magnetic field can become correlated with the electric-field polarity, leading to a systematic error. The current supply is characterized to have a stability better than 1 ppm over 100 s. The coil current of 30 mA is continuously monitored with a measurement precision of 100 pA. During the EDM measurement, the current difference between positive and negative voltages is $\Delta I = (-30\pm31)$ pA, corresponding to $\Delta B = (-2.0\pm2.1)\times10^{-3}$ pT. Using $d_B = \mu\Delta B/(2E)$ ($\mu = +0.49367(1)\ \mu_N$ [6]), the resulting false EDM is $d_B = (-6.7\pm7.0)\times10^{-29}\ e\cdot$cm. The result for this systematic effect is consistent with zero. Future experiments with improved precision on coil-current measurements would further reduce this contribution.

### C. Leakage current

A picoammeter monitors the current flowing out of the electrode to ground. We observe a difference in the leakage currents between E-field up and down, $\Delta I_{\text{leak}} = 0.24$ nA. Although the leakage current is more likely to flow along insulator surfaces, we make a conservative estimate by assuming that it flows between the electrodes, directly through vacuum along the electric-field lines, and next to the atoms. The leakage current generates loops of transverse magnetic fields. Because the electric and bias magnetic fields are misaligned by $\alpha = 22$ mrad, the loop magnetic fields have a component projected onto the bias magnetic-field direction, and the component reverses with the electric-field polarity, producing a systematic shift. The effect of this conservative estimate is $4.6\times10^{-29}\ e\cdot$cm.

### D. ODT power effect

The ODT power can influence the atomic precession frequency by a variety of high-order effects. By switching the ODT power between 10 W and 15 W, we measure a phase difference corresponding to an ODT-power dependence of $8.9\times10^{-26}\ e\cdot$cm/W. During EDM measurements, the ODT power is stabilized, and its difference between $E$-field up and down is measured to be 0.060(3) mW, yielding a systematic error below $5\times10^{-30}\ e\cdot$cm.

### E. $E^2$ effect

A residual electric-field gradient can shift the position of the atomic cloud, and affect the precession frequency when combined with a residual magnetic-field gradient. This effect scales with the square of the applied voltage. By switching the voltage between 0 and 12 kV, we measure a coefficient of $1.6\times10^{-27}\ e\cdot$cm/kV$^2$. During EDM measurements, the positive and negative voltages are matched in magnitude, and the measured difference is $0.33(31)\times10^{-3}$ kV$^2$, leading to an E$^2$ systematic error below $1\times10^{-30}\ e\cdot$cm.

## VI. MAGNETIC-FIELD NOISE

Residual circular polarization in an ODT induces a vector light shift, corresponding to a fictitious magnetic field [7],

$$\vec{B}_{\text{fict}} = \frac{A\alpha_v I}{4\epsilon_0 c g_F \mu_N F}\,\boldsymbol{e_k} \tag{10}$$

where $I$ is the ODT intensity, $A$ is the degree of circular polarization, and $\alpha_v$ is the dynamic vector polarizability. In the experiment, two methods are used to suppress the influence of $B_{\text{fict}}$. First, a calcite crystal purifies the linear polarization, giving $A < 10^{-3}$. Second, the direction of the total magnetic field is adjusted to be perpendicular to $\vec{e}_k$. Under this condition, $B_{\text{tot}} = \sqrt{B_0^2 + B_{\text{fict}}^2} = B_0 + B_{\text{fict}}^2/(2B_0)$, and the total field becomes insensitive to fluctuations of the fictitious magnetic field.

The EDM data show that the residual magnetic-field noise has been reduced to a standard deviation of 1 pT at an averaging time of 100 s, approaching the Johnson-noise-limited level [8].

## VII. THE NUCLEAR MODEL CALCULATION

In MR-CDFT, the wave functions of low-lying states in odd-mass nuclei are constructed as superpositions of quantum-number projected configurations characterized by different intrinsic quadrupole–octupole deformation parameters, $\beta_2$ and $\beta_3$, and intrinsic quantum numbers $K$, where $K$ corresponds to the projection of nuclear total angular momentum onto the body-fixed symmetric axis [9]. The potential energy of each quantum-number projected configuration with spin parity $J^\pi$ as a function of $\beta_2$ and $\beta_3$ is shown in Fig. S5. For the configurations with $J^\pi = 1/2^-$, the potential energy surface (PES) exhibits a minimum at $(\beta_2, \beta_3) = (0.38, 0.04)$, whereas for those with $J^\pi = 1/2^+$, the energy minimum is located at $(0.38, 0.12)$. The PESs are stiff along the $\beta_2$ direction but soft along $\beta_3$, indicating a stable quadrupole shape with pronounced octupole fluctuations.

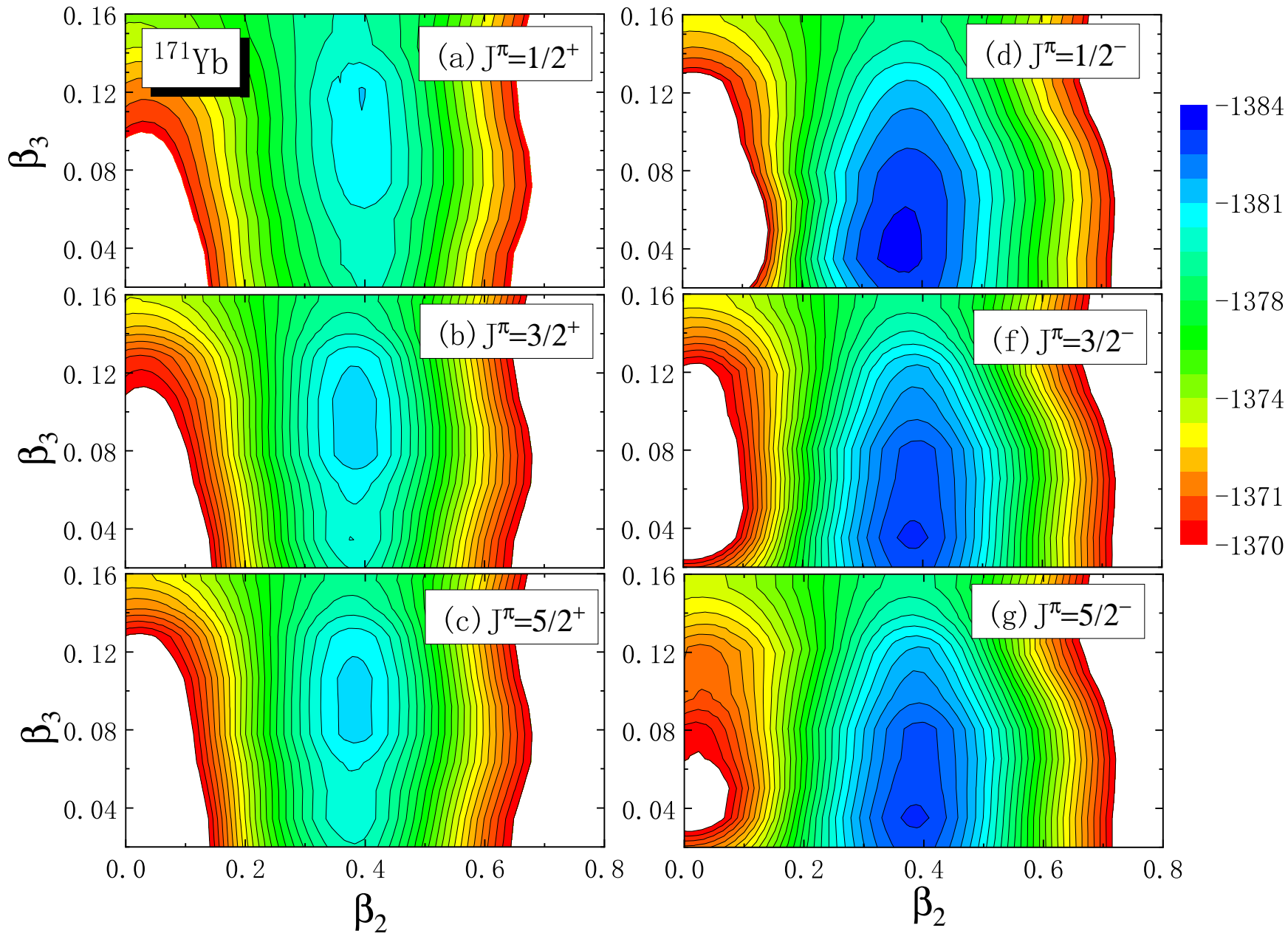

Figure S5. The potential energy surfaces of $^{171}$Yb for the configurations with spin–parity $J^\pi = 1/2^\pm$, $3/2^\pm$, and $5/2^\pm$ as functions of the quadrupole ($\beta_2$) and octupole ($\beta_3$) deformation parameters. Each panel corresponds to a different $J^\pi$ configuration, illustrating the evolution of quadrupole–octupole correlations between positive- and negative-parity states with $J$.

The admixture of the configurations in the $(\beta_2, \beta_3)$ deformation plane forms correlated wave functions for nuclear low-lying states. The energies and mixing weights are determined by the variational principle. Once the weight functions are obtained, nuclear electromagnetic observables can be evaluated using the correlated wave functions [10]. The resulting energy spectra are displayed in Fig. S6, which shows several low-lying rotational bands without corresponding nearly–degenerate parity-partner states. The ground state is correctly predicted to be $J^\pi = 1/2^-$, while the first $1/2^+$ excitation is predicted lies around 1.65 MeV. The MR-CDFT reproduces the key spectroscopic features of $^{171}$Yb, including electric quadrupole and magnetic dipole transition strengths, as shown in Fig. S7. Furthermore, the ground-state magnetic dipole moment calculated using MR-CDFT is 0.51 $\mu_N$, in agreement with the experimental value of $0.49367(1)\mu_N$ [6]. These results demonstrate that MR-CDFT provides a valid description of the low-energy structure of $^{171}$Yb.

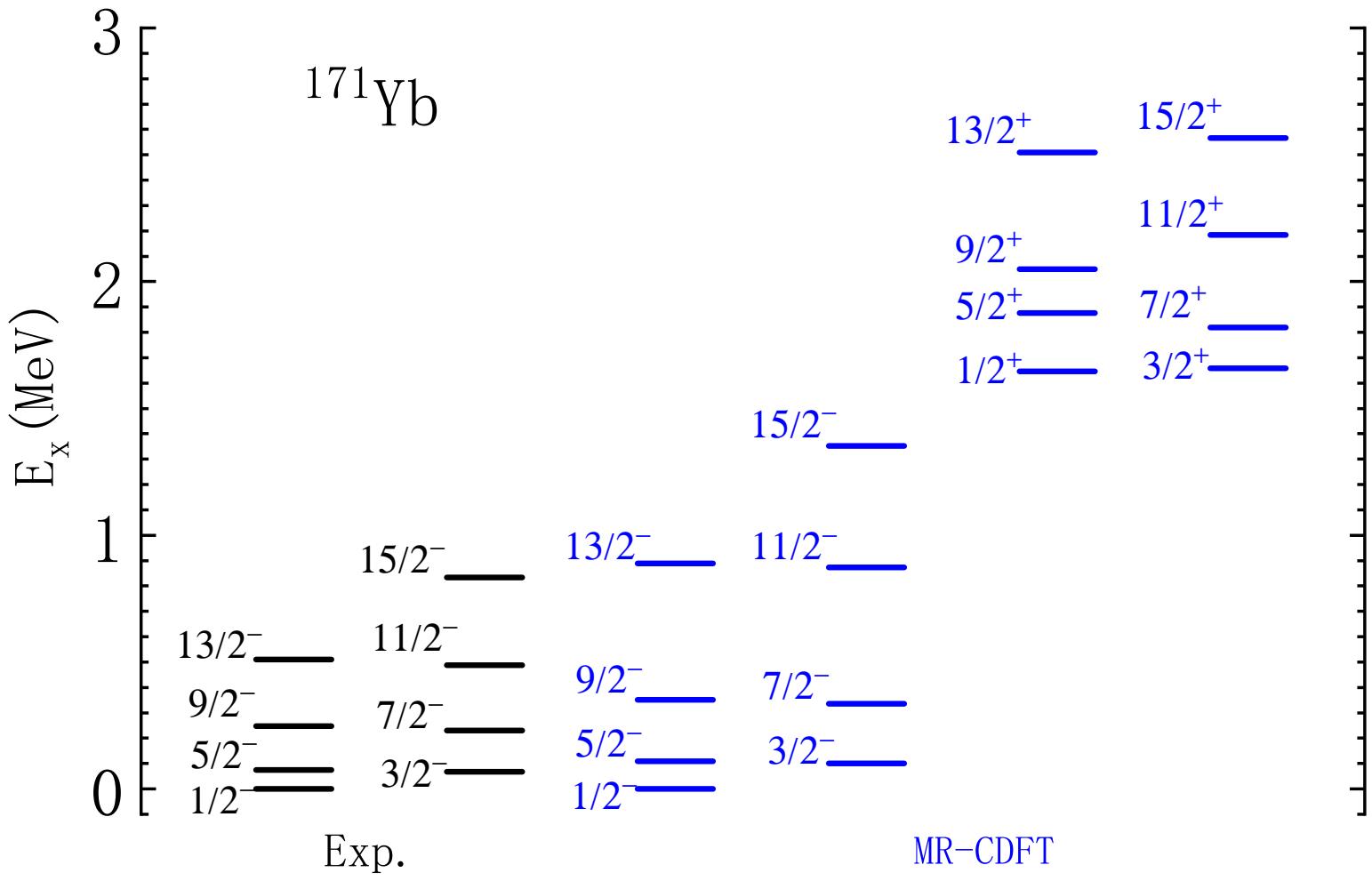

Figure S6. Calculated and measured low-lying $K = 1/2$ band energy spectra in $^{171}$Yb. Theoretical levels obtained from the MR-CDFT calculations (blue) are compared with experimental data (black) [11]. It should be noted that no experimental data are available for the positive-parity states in the $K = 1/2$ band. The overall agreement demonstrates that the MR-CDFT framework accurately captures the structure and parity splitting of the low-lying states in this odd-mass nucleus.

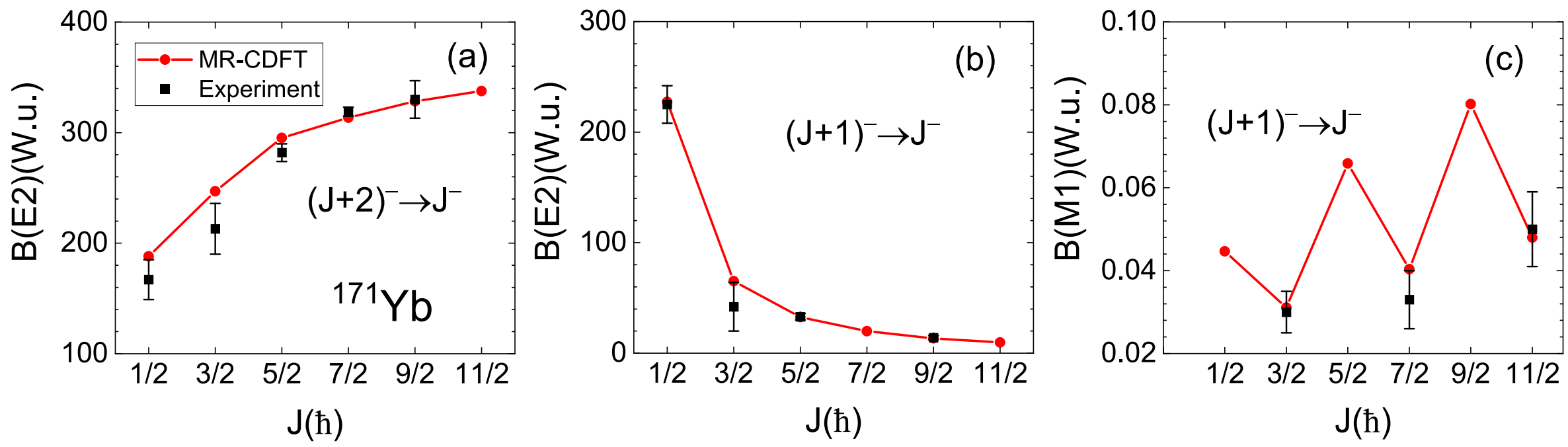

Figure S7. Comparison between the calculated and measured electromagnetic properties of $^{171}$Yb. Panels (a,b) show the electric quadrupole transition strengths $B(E2)$ for the $(J+2)^- \to J^-$ and $(J+1)^- \to J^-$ transitions in the Weisskopf units (W.u.), respectively. Panel (c) displays the magnetic dipole transition strengths $B(M1)$ for $(J+1)^- \to J^-$. Experimental data (black) are from Ref. [11].

Table III. Structure factors $a_{0,1,2}$ for $^{129}$Xe, $^{171}$Yb, $^{225}$Ra, $^{205}$Tl, and $^{199}$Hg from the MR-CDFT calculation. The results of $^{129}$Xe, $^{199}$Hg and $^{225}$Ra are taken from Ref. [12].

| **Nucleus** | $a_0(e\cdot\text{fm}^3)$ | $a_1(e\cdot\text{fm}^3)$ | $a_2(e\cdot\text{fm}^3)$ |
|---|---|---|---|
| $^{171}$Yb | -0.0053 | 0.0126 | -0.0168 |
| $^{199}$Hg | -0.0163 | 0.0327 | -0.0375 |
| $^{129}$Xe | -0.0052 | 0.0134 | -0.0097 |
| $^{225}$Ra | 0.1500 | -1.9000 | 1.3000 |
| $^{205}$Tl | 0.0110 | 0.0760 | 0.0232 |